\documentclass[a4paper,twocolumn,11pt,accepted=2025-06-10]{quantumarticle}
\pdfoutput=1
\usepackage[utf8]{inputenc}
\usepackage[english]{babel}
\usepackage[T1]{fontenc}
\usepackage{amsmath}
\usepackage{hyperref}
\usepackage{amsfonts}
\usepackage{dsfont}

\usepackage{tikz}
\usepackage{lipsum}

\renewcommand{\vec}[1]{\boldsymbol{#1}} 

\begin{document}

\title{Deep Circuit Compression for Quantum Dynamics via Tensor Networks}

\author{Joe Gibbs}
\affiliation{School of Mathematics and Physics, University of Surrey, Guildford, GU2 7XH, UK}
\affiliation{AWE, Aldermaston, Reading, RG7 4PR, UK}

\author{Lukasz Cincio}
\affiliation{Theoretical Division, Los Alamos National Laboratory, Los Alamos, NM 87545, USA}

\begin{abstract}
Dynamic quantum simulation is a leading application for achieving quantum advantage. However, high circuit depths remain a limiting factor on near-term quantum hardware. We present a compilation algorithm based on Matrix Product Operators for generating compressed circuits enabling real-time simulation on digital quantum computers, that for a given depth are more accurate than all Trotterizations of the same depth. By the efficient use of environment tensors, the algorithm is scalable in depth far beyond prior work, and we present circuit compilations of up to 64 layers of $SU(4)$ gates.
Surpassing only 1D circuits, our approach can flexibly target a particular quasi-2D
gate topology. We demonstrate this by compiling a 52-qubit 2D Transverse-Field Ising propagator onto the IBM Heavy-Hex topology. For all circuit depths and widths tested, we produce circuits with smaller errors than all equivalent depth Trotter unitaries, corresponding to reductions in error by up to 4 orders of magnitude and circuit depth compressions with a factor of over 6.
\end{abstract}

\maketitle

\section{Introduction}

The dynamic simulation of quantum many-body systems was an initial motivation for the use of quantum computers~\cite{feynman1982simulating}, and is still considered a leading application for achieving quantum advantage~\cite{kim2023evidence}.
This can be attributed to a range of factors, including a compact mapping onto qubits,
a repeating circuit structure aiding noise model characterization~\cite{berg2022probabilistic}, and a rapid growth in entanglement limiting classical simulations~\cite{schollwock2011density}. 

\begin{figure}[b!]
\centering
\vspace{-5mm}
\includegraphics[width=0.95\columnwidth]{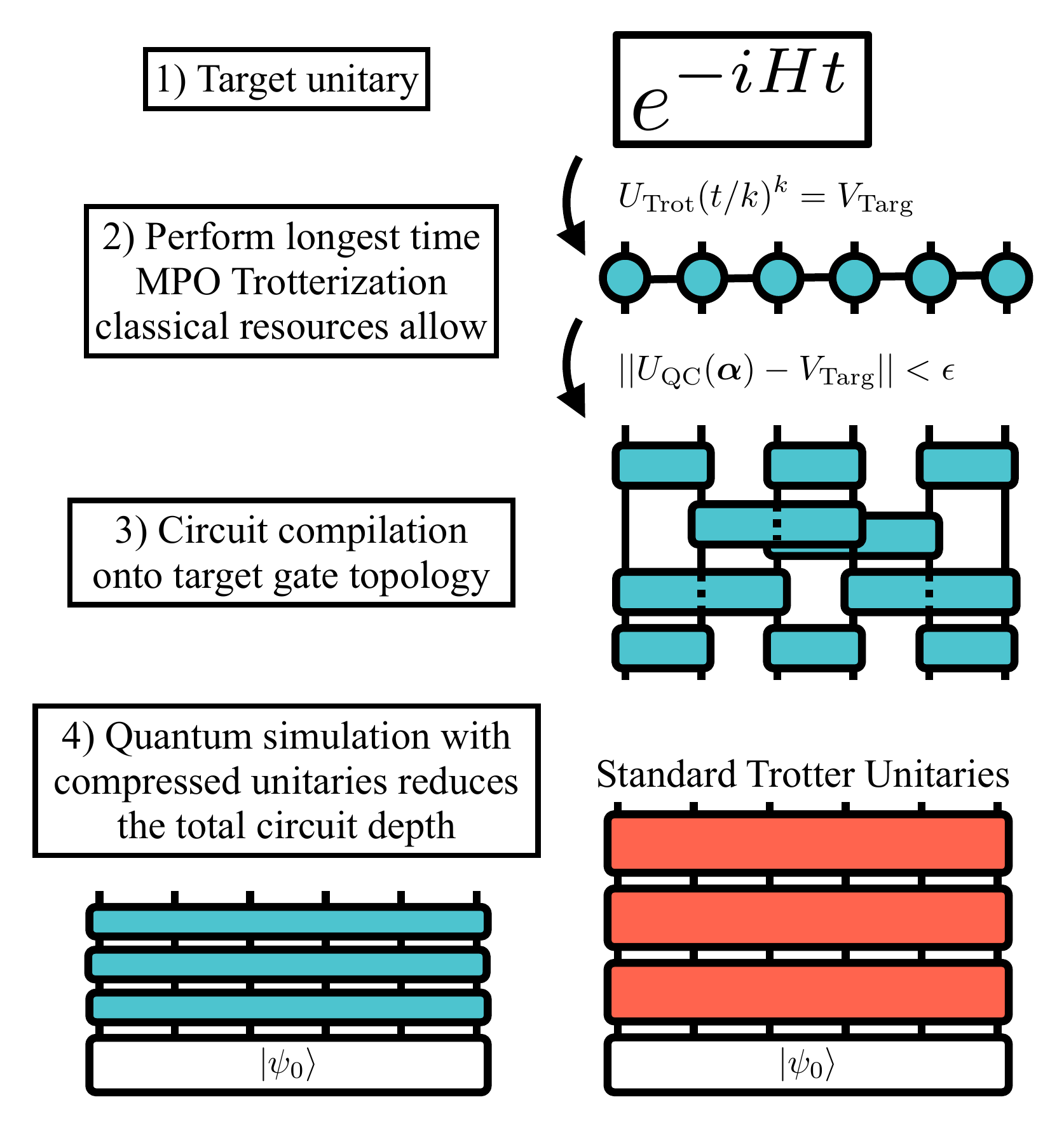}
\vspace{-3mm}
\caption{
\textbf{Overview of Algorithm. } 
1) The input of the algorithm is a Hamiltonian $H$ and time step $t$. The algorithm finds a shallow circuit that approximates the propagator $e^{-iHt}$.
2) The propagator is represented as an MPO by Trotterization with a fine time step for a negligible Trotter error, for the longest time that results in an MPO with a tractable bond dimension.
3) For the gate topology of a target quantum computer, an optimization is performed to maximize the overlap between the target unitary $V_{\text{Targ}}$ and the variational quantum circuit $U_{\text{QC}}(\vec{\alpha})$.
4) The compressed circuit for the propagator can enable real-time simulation on a quantum computer with a shorter circuit depth than possible using standard Trotterization methods.
}
\label{fig:Cartoon}
\end{figure}

Currently, the most popular approach for performing real-time simulation of quantum dynamics is through the use of product formulas~\cite{lloyd1996universal}. The propagator $e^{-iHt}$ for a Hamiltonian $H$ cannot be directly applied on digital quantum computers, so it is approximated via Trotterization into time steps composed of lower body operators. Trotter decompositions come in a range of orders that vary in circuit depth and associated error scalings~\cite{childs2021theory}. To reach a tolerable simulation error with product formulas, large scale experiments  are currently limited by the deep required circuit depths beyond the coherence time of NISQ devices available quantum hardware.

The long term goal of quantum simulation will be enabled by error-corrected fault-tolerant quantum computers. Today's noisy intermediate scale quantum devices (NISQ)~\cite{preskill2018quantum} are however characterized by an intermediate number of physical qubits with restrictions on circuit depth due to hardware noise. Reducing the number of noisy operations in a quantum circuit, notably the 2-qubit gate count, is critical for achieving a near-term quantum advantage, and remains the primary concern for NISQ algorithm designers.

The need for automated methods of generating short-depth circuits motivated the development of Variational Quantum Algorithms (VQAs)~\cite{cerezo2020variationalreview}. These are hybrid quantum-classical algorithms that construct a cost function to measure and improve the quality of a parameterized quantum circuit for performing a given task, while constraining the overall quantum computational resources used.

VQAs have been developed to allow a search for compressed circuits generating a target unitary transformation~\cite{berthusen2022quantum, mizuta2022local, bilkis2021semi}. Khatri \emph{et al} ~\cite{khatri2019quantum} proposed the Hilbert-Schmidt Test in their Quantum-Assisted Quantum Compiling (QAQC) algorithm, a circuit to allow the evaluation of the global overlap between two unitaries with a fixed depth overhead. Given a target unitary, QAQC can then variationally learn an equivalent unitary with smaller gate counts, and compile to a particular quantum computers gateset. Following work used novel out-of-distribution generalization properties of quantum neural networks~\cite{caro2021generalization, caro2022outofdistribution,gibbs2024dynamical} to allow unitary compilation with significantly reduced circuit resources. VQAs however have well known trainability issues, primarily caused by the infamous Barren Plateau phenomenon~\cite{mcclean2018barren, larocca2024review}, and complex cost function landscapes with many local minima~\cite{anschuetz2022quantum}, limiting large-scale implementations so far. Motivated by the numerous challenges of NISQ computations, researchers seek to maximally utilize the advanced classical computers available to support the quantum computer. 

Tensor networks~\cite{orus2014practical} are the leading framework for simulating quantum circuits~\cite{pan2022simulation, tindall2023efficient}, and we use Matrix Product Operators (MPO)~\cite{mcculloch2007density} to express and compress our circuits. Prior work has investigated the classical compilation of quantum states represented as Matrix Product States (MPS), into a quantum circuit to be uploaded onto the quantum computer ready for further study~\cite{rudolph2022decomposition, rudolph2022synergy, dborin2022matrix, jamet2023anderson, rogerson2024quantum, anselme2024combining, lin2021real, robertson2024tensor}.
Specifically, we target the common task of performing dynamic Hamiltonian simulation. Refs.~\cite{mansuroglu2023variational, kotil2024riemannian} proposed variational algorithms for classical optimization of translationally-invariant circuits to reduce errors compared to equivalent depth Trotterizations of the evolution operator.
Recent work has also represented quantum circuits as Matrix Product Operators to classically find short-depth unitaries for approximating Trotter evolution~\cite{causer2023scalable,mc2023classically, robertson2024tensor,anselme2024combining} . 

Recent work has explored the use of MPO-based compilations to compress the time evolution operator $e^{-iHt}$. Causer \emph{et al}~\cite{causer2023scalable} provided a local optimization based on the polar decomposition, however the method presented for contracting gate environments had an exponential scaling in the number of layers. McKeever and Lubasch~\cite{mc2023classically,mc2024towards} similarly demonstrated unitary compilations for shallow circuit depths. Tepaske \emph{et al}~\cite{tepaske2023optimal,tepaske2024optimal} gave an MPO-based compilation for this task with simulations on systems sizes of up to 16 qubits.

Our work extends the literature by enabling compilation of large scale, deep circuits, with flexibility to target arbitrary topology NISQ devices. Our algorithm allows deeper circuits to be optimized than typically expected to be representable by an MPO, because the gates in our circuit are primarily used to disentangle the target unitary. We demonstrate compressions on large scale and deep parameterized circuits, on up to 200 qubits and 64 layers of $SU(4)$ gates.

An overview of our algorithm is given in Fig.~\ref{fig:Cartoon}. For a target Hamiltonian, $H$, we compute the propagator $e^{-iHt}$ for the longest time the unitary can be represented as an MPO with a tractable bond dimension. Then we perform a classical circuit compilation onto the gate topology of a target quantum computer, where an optimization of the gates minimizes the error compared to the target time evolution unitary. The optimization of the parameterized quantum circuit is always able to reduce the approximation error to the target propagator compared to equivalent depth Trotter unitaries. Then, this compressed unitary may be uploaded to the quantum computer and applied multiple times, enabling a real-time simulation of quantum dynamics with a shorter depth than possible with standard Trotterization methods. 

After initializing the quantum circuit with an equivalent depth Trotterization, we are always able to find a reduction in the approximation error while maintaining the same circuit depth.  For the range of Hamiltonians we test on, including a quasi-2D Hamiltonian expressed on the IBM Heavy-Hex gate topology, we show that our compressions result in up to 4 orders of magnitude reduction in the error, and depth reductions by over a factor of 6.

The rest of the paper is organized as follows. Section~\ref{sec:methods} introduces the error metric used to compare approximation errors of unitaries, the efficient use of environment tensors, and the full optimization procedure used. Section~\ref{sec:results} presents the results of our compression algorithm applied to a range of Hamiltonians, and gives numerical evidence supporting our ability to optimize deep circuits. Finally, Section~\ref{sec:discussion} gives a discussion of our work.

\section{Methods}\label{sec:methods}

\subsection{Error Metric}

To evaluate the performance of our compilations, we require a measure of distance between $n$-qubit unitaries. The metric we use is the Hilbert-Schmidt Test (HST), given by 
\begin{equation}
    C_{\text{HST}}(U, V) = 1-\frac{1}{2^{2n}}|\text{Tr}(U^{\dagger}V)|^2.
\end{equation}

This is an operationally meaningful quantity as it is directly related to the average state infidelity when applying the two unitaries to a random state~\cite{caro2022outofdistribution} e.g.

\begin{equation}
    C_{\text{HST}}(U, V) = \frac{2^n + 1}{2^n}(1 - \mathbb{E}_{|\psi\rangle \sim \mathcal{S}_{\text{Haar}_n}} |\langle\psi|U^{\dagger}V|\psi\rangle|^2),   
\end{equation}
where $\mathcal{S}_{\text{Haar}_n}$ is the $n$-qubit Haar random distribution.
For our compilations, we have the target unitary $V_{\text{Targ}}$ and a variational quantum circuit $U_{\text{QC}}(\vec{\alpha})$, with these a more explicit version of the cost function we optimize over reads

\begin{equation}\label{eq:HSTcost}
    C_{\text{HST}}(\vec{\alpha}) = 1-\frac{1}{2^{2n}}|\text{Tr}(U_{\text{QC}}(\vec{\alpha})^{\dagger}V_{\text{Targ}})|^2.
\end{equation}

$C_{\text{HST}}(\vec{\alpha})$ could be directly computed from the inner product $\text{Tr}(U_{\text{QC}}(\vec{\alpha})^{\dagger}V_{\text{Targ}})$, however for $n$-qubit operators this is an exponentially large quantity. To avoid potential floating-point overflow errors, we instead convert our MPOs into MPS on a doubled physical space, normalize, and then take their inner product. This allows the HST to be computed faithfully, without needing to store exponentially large numerical values. The normalization of this MPS is equivalent to rescaling the original MPO to have a norm of $2^N$ as expected for a unitary. If this step was not performed, our cost function may become unfaithful and record negative values from to the deviation of the norm of the MPO due to truncation errors.


\subsection{Constructing the target MPO}\label{sec:CreateMPO}

Our algorithm begins by constructing the propagator $e^{-iHt}$ for a target Hamiltonian $H$, represented as an MPO. The bond dimension of this operator is governed by the total evolution time, the type of interactions, and their range when mapped onto a 1D chain of qubits.

The propagator is constructed by decomposing the operator into a very deep circuit using a high order Trotterization, which is then contracted into an MPO. Specifically we use 
\begin{equation}\label{eq:4thOrder}
    e^{-iHt} \approx U_{4\text{th}}(t/k)^k,
\end{equation}
where $U_{4\text{th}}(t)$ is a 4th-order Trotterization, and we set $k=10$ which has a negligible Trotter error for the Hamiltonians and time steps we study.

\begin{figure*}[t!]
\centering
\includegraphics[width=0.9\textwidth]{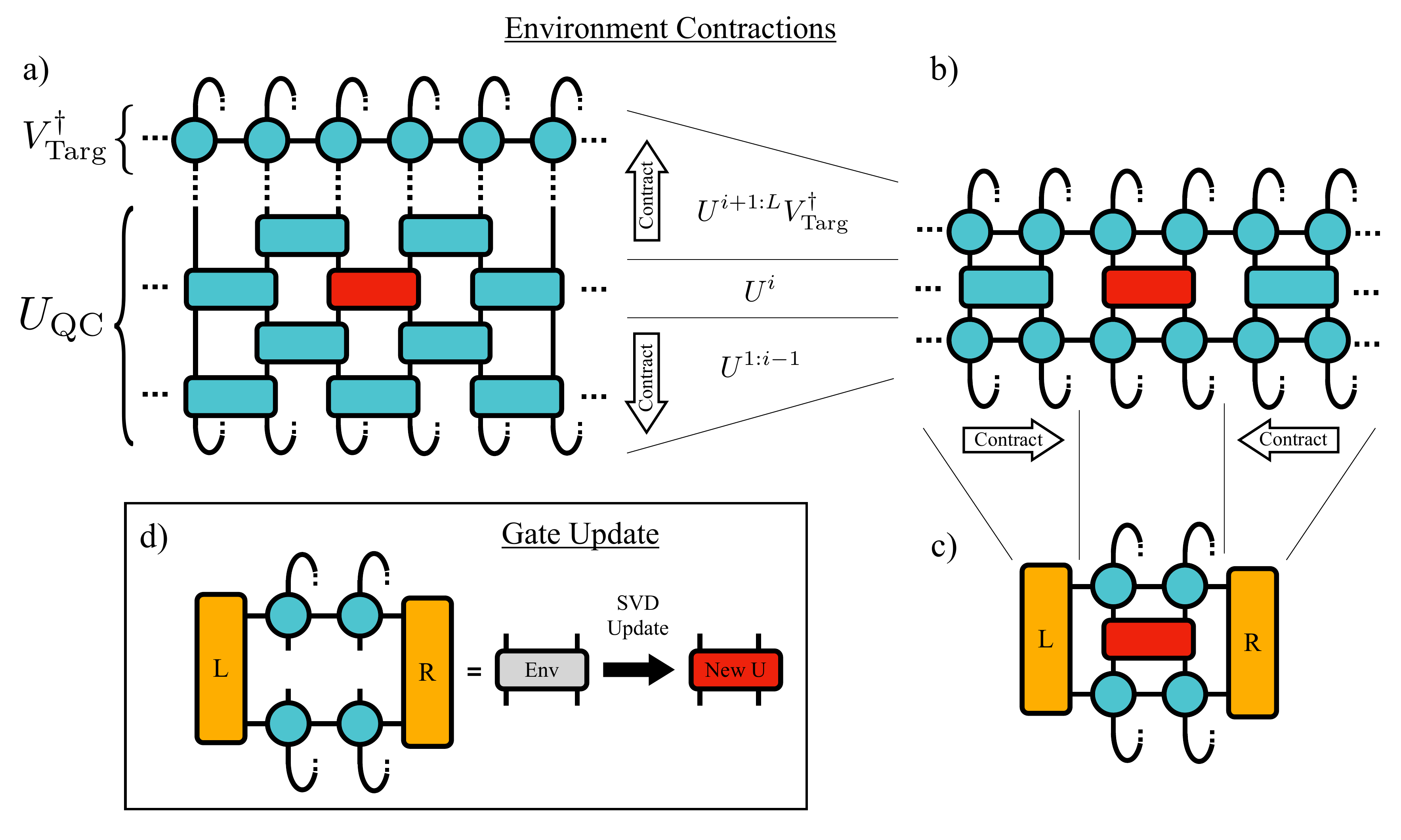}
\vspace{-2mm}
\caption{\textbf{Efficiencies in Environment Contractions. } a) The contraction for the inner product of the quantum circuit $U_{\text{QC}}$ and the target unitary $V_\mathrm{targ}$. We iterate over the gates, updating their state to increase the inner product. Starting at the top layer of the circuit, we move down to the first layer and back up to the top. This is one `sweep'. The curved lines indicate the top and bottom physical legs are connected. b) When optimizing a particular layer of the circuit, the tensor network outside the row of gates is unchanged. To avoid unnecessary re-contraction, we store the state of the network above and below the current layer with environment MPOs. 
c) When moving along one gate in the layer, the network is fixed to the left and the right. The state outside the current gate is represented by contracted left and right environment tensors. We can therefore cheaply move back and forth updating gates in the layer by reusing the environment tensors $L$ and $R$.
d) The environment of the gate is computed, which is SVD'd to find the locally optimal gate and updated in the circuit.}
\label{fig:Contract}
\end{figure*}

We seek to compress unitaries enabling dynamic simulation for the longest time where the target MPO is still representable by an MPO with tractable bond dimension. To determine this longest time, we do a parameter sweep across a range of timescales. For each total evolution time, we iteratively perform the MPO Trotterization for an increasing maximum bond dimension, and compute $C_{\text{HST}}$ between the consecutive MPOs. We consider that quantity to be converged when it is below $10^{-10}$. We then select the propagator to train on as the largest total evolution time where the iterative MPO Trotterization converged below a final bond dimension threshold. We set this bond dimension to be 128 for our numerical results. 

When performing time evolution on a NISQ device, compiling the propagator with an arbitrarily small truncation error will be redundant compared to the compilation error of the outputted quantum circuit, and errors due to hardware noise. For example, if our final compilation error is 
$C_{\text{HST}}(U_{\text{QC}},  V_{\text{Targ}}) \approx 10^{-4}$ , then our total truncation error does not need to be lower than $C_{\text{HST}}(V_{\text{Targ}}, e^{-iHt}) \approx 10^{-5}$.
Therefore, before beginning the optimization, we compress the MPO to an error threshold below the target compilation error, which can significantly reduce the bond dimension of the target MPO and the computational burden during training. As a result, this observation enables us to extend the simulation time that is attainable with our method.

Throughout our work we denote this compressed MPO as $V_{\text{Targ}}$. The MPO compression procedure used is a variational compression~\cite{schollwock2011density}. The MPO is transformed to an MPS on a doubled physical space, truncated to a lower bond dimension, and then variationally optimized until the fidelity with the uncompressed MPS has converged, and then remapped back into an MPO. We use a single singular value decomposition (SVD) compression sweep as an initialization. If the cost $C_{\text{HST}}$ between the target and variationally compressed MPO is above the target error fidelity, we repeat this procedure for a higher bond dimension until the desired error condition is satisfied.

\begin{figure*}[t!]
\centering
\includegraphics[width=0.95\textwidth]{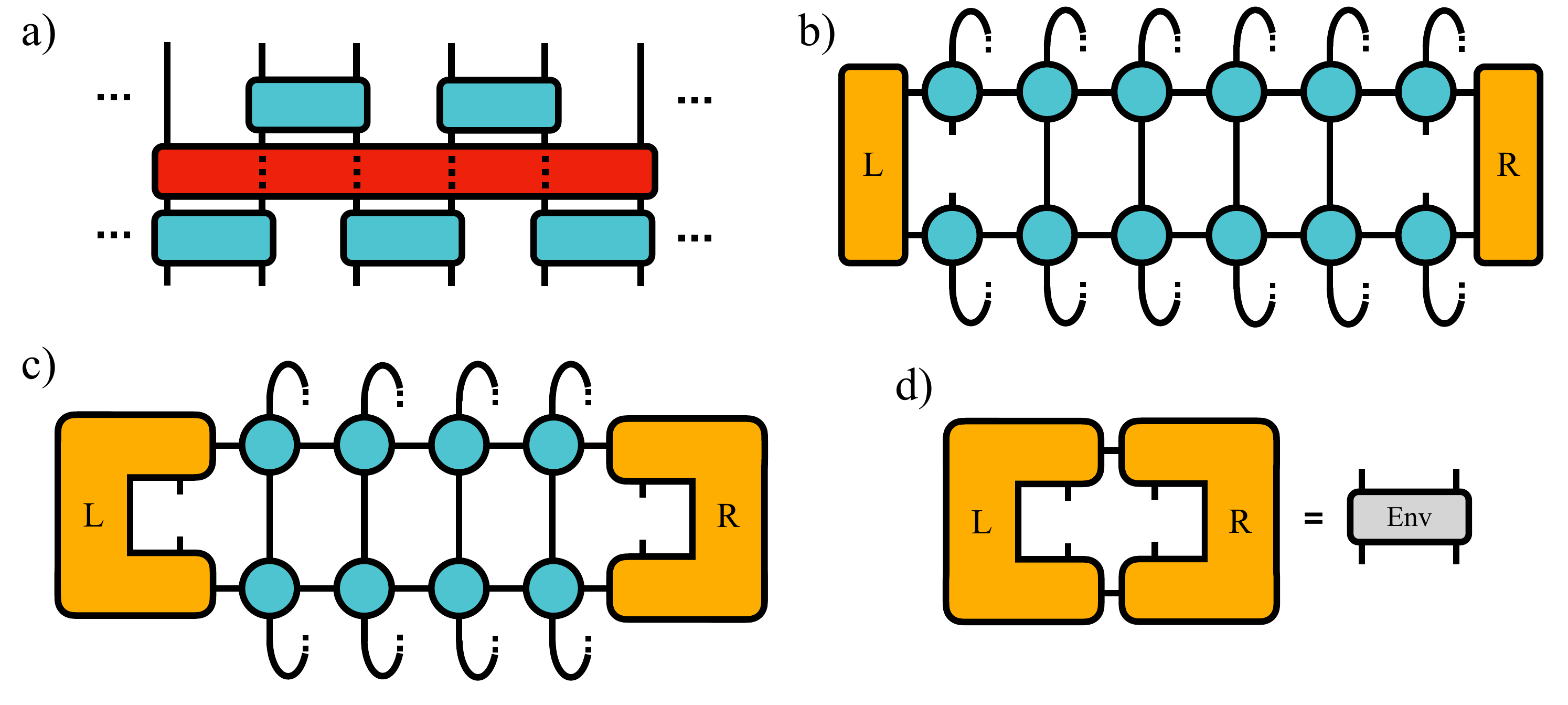}
\vspace{-2mm}
\caption{\textbf{Environments of non-local gates. } a) A flexible topology circuit compilation may involve local 2D gates mapping onto non-local 1D gates. b) The tensor network giving the environment for a non-local gate acting on qubit indices $\{i, i+5\}$.  c) The contraction can maintain a $\mathcal{O}(\chi^2)$ memory and $\mathcal{O}(\chi^3)$ runtime by iteratively contracting neighboring MPO tensors into the left/right environment tensors. d) After all intermediate MPO site tensors have been absorbed, the left and right environment tensors are contracted to give the final gate environment. }
\label{fig:LRContract}
\vspace{-1.5em}
\end{figure*}

\subsection{Optimization}

We use an optimization scheme where all gates are iteratively swept through and updated to reduce the cost function given in Eq.~\eqref{eq:HSTcost}. For each individual gate, we find their environment in the tensor network, which is singular value decomposed to find the optimal unitary to maximize the overlap.  This is a standard optimization procedure in tensor network algorithms~\cite{evenbly2009algorithms}, and has been used for the classical optimization of circuits represented by state vectors~\cite{kukliansky2023qfactor, shirakawa2021automatic, kukliansky2024leveraging},  MPS~\cite{rudolph2022decomposition, anselme2024combining} and MPO~\cite{causer2023scalable, anselme2024combining}.

To minimize the total circuit depth, we choose our ansatz structure to be the most dense packing of commuting layers of $SU(4)$ gates, which by the KAK decomposition are known to require at most 3 entangling gates~\cite{tucci2005introduction}. Our circuits can be tailored to the specific gate topology of the target quantum computer, rather than assuming access to a 1D chain of qubits with nearest-neighbor connectivity. We detail this further in Section~\ref{sec:2D}.

A circuit composed of $L$ layers of commuting gates can be denoted $U^{1:L} = \prod_{i=1}^L U^i$, where $U^i$ is the $i^{\text{th}}$ layer of commuting gates. At the $i^{\text{th}}$ layer we optimize over the gates contained in $U^{i}$ to minimize $C_{\text{HST}}(\vec{\alpha})$.
We update the circuit in ``sweeps'', where we start at $U^L$, move down to $U^1$, and back up to $U^L$. 

The network is contracted into top/bottom environment MPOs representing $U^{1:i-1}$ and $U^{i+1:L} V_\mathrm{Targ}^{\dagger}$ respectively, see Fig.~\ref{fig:Contract}b). For a deep and wide circuit, contracting these MPOs is a costly operation. Here we take advantage of the update order to more efficiently reuse and update these environments. When moving sequentially through the circuit, the tensor network is only updated locally within a layer. When moving to the neighboring layer, we can cheaply update the environment MPOs by absorbing the new layer of gates, rather than recontracting the whole network. 
All circuit templates in this work are derived from Trotterizations, and consecutive layers within the circuits do not commute. However the gates within a particular layer do mutually commute, even for the circuits compiled onto a 2D topology, as each gate acts on disjoint qubits.

When optimizing a particular gate in a layer, we find its local environment by contracting the tensor network shown in Fig.~\ref{fig:Contract}b) from the left and the right. This gives the gate's environment tensor by the contraction shown in Fig.~\ref{fig:Contract}d). With these left/right environment tensors, denoted by $L$/$R$ in Fig.~\ref{fig:Contract}c), we can cheaply move left/right optimizing the gates within the layer multiple times by locally updating the environment tensors.

While the computation of the required quantities can be made efficient, the navigation during the optimization over the manifold of unitaries can remain challenging, and a good initialization is critical. We find that for a target circuit depth, initializing with an equivalent depth Trotterization gives excellent results with immediate decreases in the cost function, and we use this initialization technique to obtain all our results described in Sec.~\ref{sec:results}.

\subsection{MPO Resets}

We note that over the course of many optimization sweeps, truncation errors will cause the MPO representation of the tensor network environment above and below the current row to deviate from faithfully representing the current state of the quantum circuit, causing the computation of incorrect gate environments and degrading the optimization. However, consider the environment MPOs for the outer layers $U^1$ and $U^L$. For $U^1$, the top and bottom environment MPOs represent $U^{2:L} V_{\text{Targ}}^{\dagger}$ and $\mathds{1}$ (the identity) respectively, and for $U^L$ the top and bottom environment MPOs represent $V_{\text{Targ}}^{\dagger}$ and $U^{1:L-1}$. In both of these cases, one of the environment MPOs represents a unitary that remains fixed throughout the optimization, $\mathds{1}$ and $V_{\text{Targ}}^{\dagger}$. As a result, at these outer layers one of the environment MPOs can be reset to their exact value, allowing a simple and cheap method for entirely removing these accumulating truncation errors.

\begin{figure*}[t!]
\centering
\includegraphics[width=0.9\textwidth]{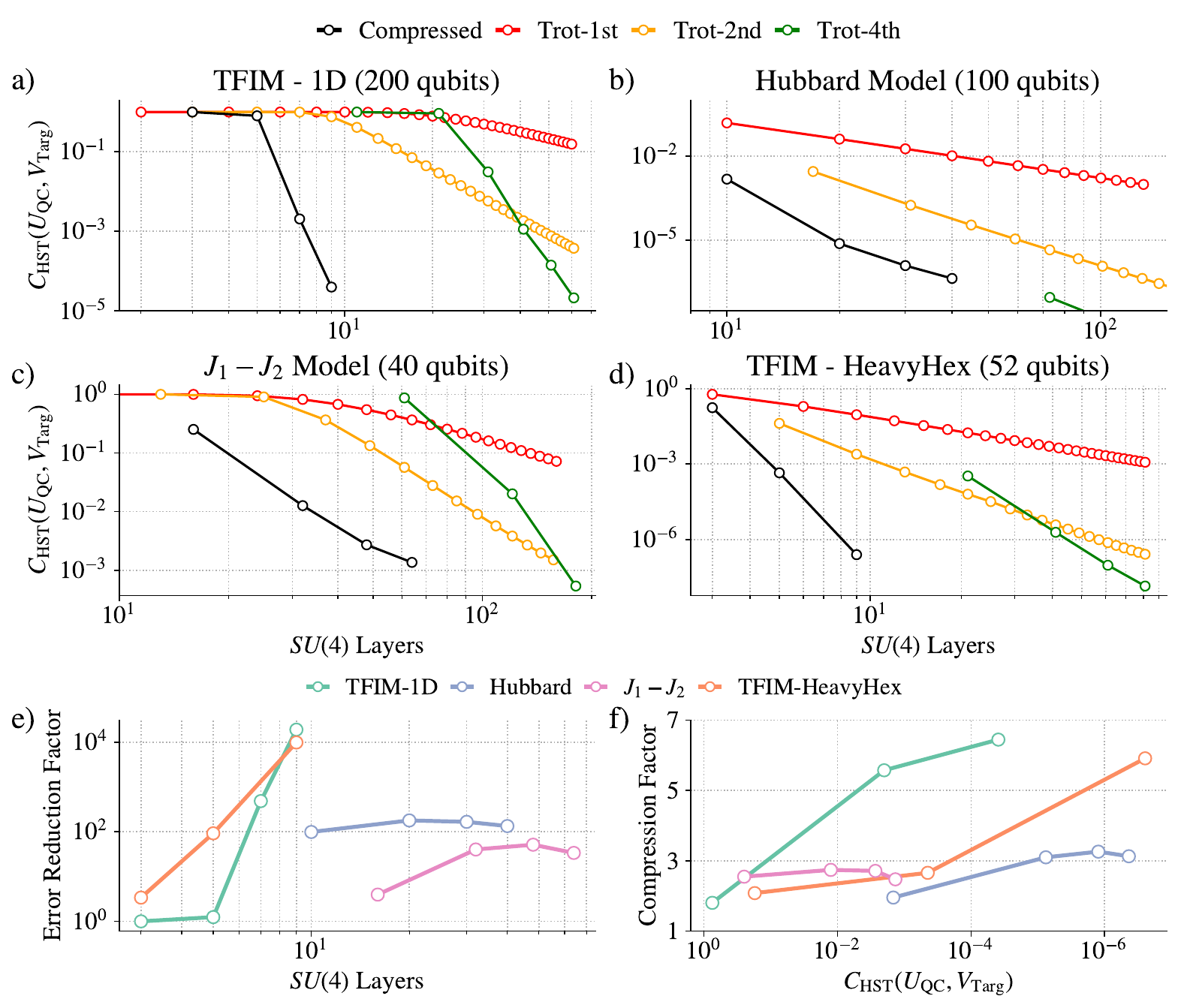}
\vspace{-2mm}
\caption{\textbf{Compression of Trotter Unitaries } a)-d) For a range of target Hamiltonians, we compile circuits to minimize the approximation error with the propagator for a range of Hamiltonians, specified in Appendix~\ref{sec:Hamiltonians}. The final error upon completion of the optimization is measured in comparison to increasing depth Trotterizations. We compare against Trotterized unitaries of the form $U_r(t/k)^k$ for $r$ = 1st, 2nd and 4th order Trotterizations, shown by the red, orange and green curves. Our numerical results present compressions of Trotter unitaries of up to 200 qubits, and 64 layers of commuting $SU(4)$ gates.
e) For each of the compressed circuits, we measure factor by which the cost is reduced compared to the equivalent depth Trotter unitary, with reductions presented up to 4 orders of magnitude. When comparing against Trotter, we use the interpolation of the data points. f) We also display the depth compression factor, which for a particular compressed circuit is the ratio of the circuit depth compared to the shortest Trotter unitary with the same error versus $V_{\text{Targ}}$.}
\label{fig:Numerics}
\vspace{-1.5em}
\end{figure*}

\subsection{Targeting a 2D gate topology} \label{sec:2D}

Prior works on MPO-based circuit compilations have assumed access to a 1D chain of qubits with nearest-neighbor gates. We extend this to enable a flexible compilation onto the 2D qubit connectivity of a target quantum computer. MPS are a fundamentally 1D tensor network, and a result will be exponentially impeded in one dimension as the size of a target 2D system increases. However, due to their favorable runtime scaling as a function of bond dimension, MPS have been successfully applied to model restricted-width 2D systems, e.g., on infinite cylinders, where local 2D interactions are transformed into non-local 1D interactions. For MPS simulations of a 2D lattice with dimensions $L_x \times L_y$, the computational cost is linear in $L_x$ and exponential in $L_y$. A calculation scalable in both dimensions must be based on a corresponding 2D tensor network, which is beyond the scope of this work. However, the extension to 2D systems of limited width still provides utility for hardware demonstrations on available quantum hardware, extending the literature of tensor network circuit compilations and exploring the limits of MPS-based compilation of quantum dynamics.

For the compression of unitaries on a 2D gate topology, similarly we will encounter non-local gates acting on the MPO and need an extended contraction scheme for computing gate environments. The gates present in a Trotterization of a 2D Hamiltonian will inevitably result in non-local gates when mapped onto the MPO, requiring an extended contraction scheme for computing gate environments. Fig.~\ref{fig:LRContract} shows the required contraction when a gate acts on MPO sites $\{i, i+5\}$. To maintain a scaling of $\mathcal{O}(d^2\chi^2)$ for memory and $\mathcal{O}(d^3\chi^3)$ for runtime (where $d$ is the physical site dimension, and $\chi$ is the MPO virtual bond dimension), the contraction is performed by iteratively absorbing the intermediate MPO site tensors into the left/right environment tensors and normalizing, until these are directly connected and can be contracted to give the final gate environment. As described earlier, the environment is SVD'd to find the new locally optimal gate. When finished optimizing a given circuit layer containing these non-local gates, and we wish to apply this layer into the top/bottom environment MPOs, the non-local gates are decomposed into a 2-body MPO and applied using an MPO-MPO zip-up, similar to the MPO-MPS zip-up algorithm proposed in~\cite{stoudenmire2010minimally}.

While the computation of the target MPO was not the dominant cost in our numerical results, this subroutine would become more expensive and limiting for Hamiltonians with longer-range interactions when mapped onto a chain of qubits, for example molecular systems in quantum chemistry. Our compilation only requires access to an MPO representing $e^{-iHt}$ to optimize a variational circuit against. As a result, we do not explicitly require a circuit to compress. More efficient methods of constructing high-order MPOs for time evolution~\cite{van2024efficient,zaletel2015time} could be used instead, which may extend the range of systems and simulations times our compilations can be applied to.


\section{Results}\label{sec:results}

We test our compression algorithm against a range of Hamiltonians: (i) the 1D Transverse-Field Ising model, on 200 qubits for an evolution time of $t=2.0$; (ii) the 1D Hubbard model, on 100 qubits for an evolution time of $t=0.2$; (iii) the 1D $J_1-J_2$ model, on 40 qubits for an evolution time of $t=1.0$; and (iv) a 2D Transverse-Field Ising model, with the same connectivity graph as the 52 qubit subset of the IBM Heavy-Hex topology shown in Fig.~\ref{fig:HeavyHex}, for an evolution time of $t=0.5$. These Hamiltonians are specified in more detail in Appendix~\ref{sec:Hamiltonians}. We highlight that for the 2D Transverse-Field Ising model demonstration, we perform the compilation assuming access to a quantum computer with the same nearest-neighbor interactions of the gate topology. When mapped onto the MPO, gates that are local on Heavy-Hex lattice will be transformed into non-local 1D gates, and we detail the extension this requires in Sec.~\ref{sec:2D}.
Our compressed circuits are tested against a range of Trotter decompositions. Specifically, for a given parameterized circuit $U_{\text{QC}}(\vec{\alpha})$, we are interested in the error $C_{\text{HST}}(U_{\text{QC}}(\vec{\alpha}), e^{-iHt})$, and we express $e^{-iHt}$ by the highly accurate Trotter decomposition given in Eq.~\eqref{eq:4thOrder}. We test this against Trotter unitaries of the form $U_r(t/k)^k$ for $r$ = 1st, 2nd and 4th order Trotterizations, shown by the red, orange and green curves in Fig.~\ref{fig:Numerics}a)-d).

In computing a Trotter unitary for a given Hamiltonian, terms can be reordered to allow neighboring layers of gates acting on the same qubit indices to be absorbed into a single layer. We maximally exploit this to compare our compressed circuits to the shortest depth comparable Trotterization.

For all circuit depths tested, our compressed circuits are able to achieve a smaller approximation error to the target unitary compared to all equivalent depth Trotterizations. During the optimization, we train with a constant maximum bond dimension of 128. However to faithfully determine the progress of the optimization, all data points for  shown in Fig.~\ref{fig:Numerics} has been obtained by testing $C_{\text{HST}}(U_{\text{QC}}, V_{\text{Targ}})$ with a doubled bond dimension when computing the MPOs $U_{\text{QC}}$ and $V_{\text{Targ}}$. 

For all Hamiltonians tested, we find at least a factor of 10 reduction in the cost compared to the best Trotterization, and up to a factor of $10^4$ for the Transverse-Field Ising model. Conversely, to achieve a particular compilation error compared to the target unitary, our compressed circuits are typically at least 2 times shallower than the Trotterization with the same error, and up to 6.4 times shallower. We are generous to Trotterization by comparing against the polynomial interpolation of the data points, even if a Trotterization does not exist at that exact depth. 

\begin{figure}[t!]
\centering
\vspace{+1.5em}
\includegraphics[width=\columnwidth]{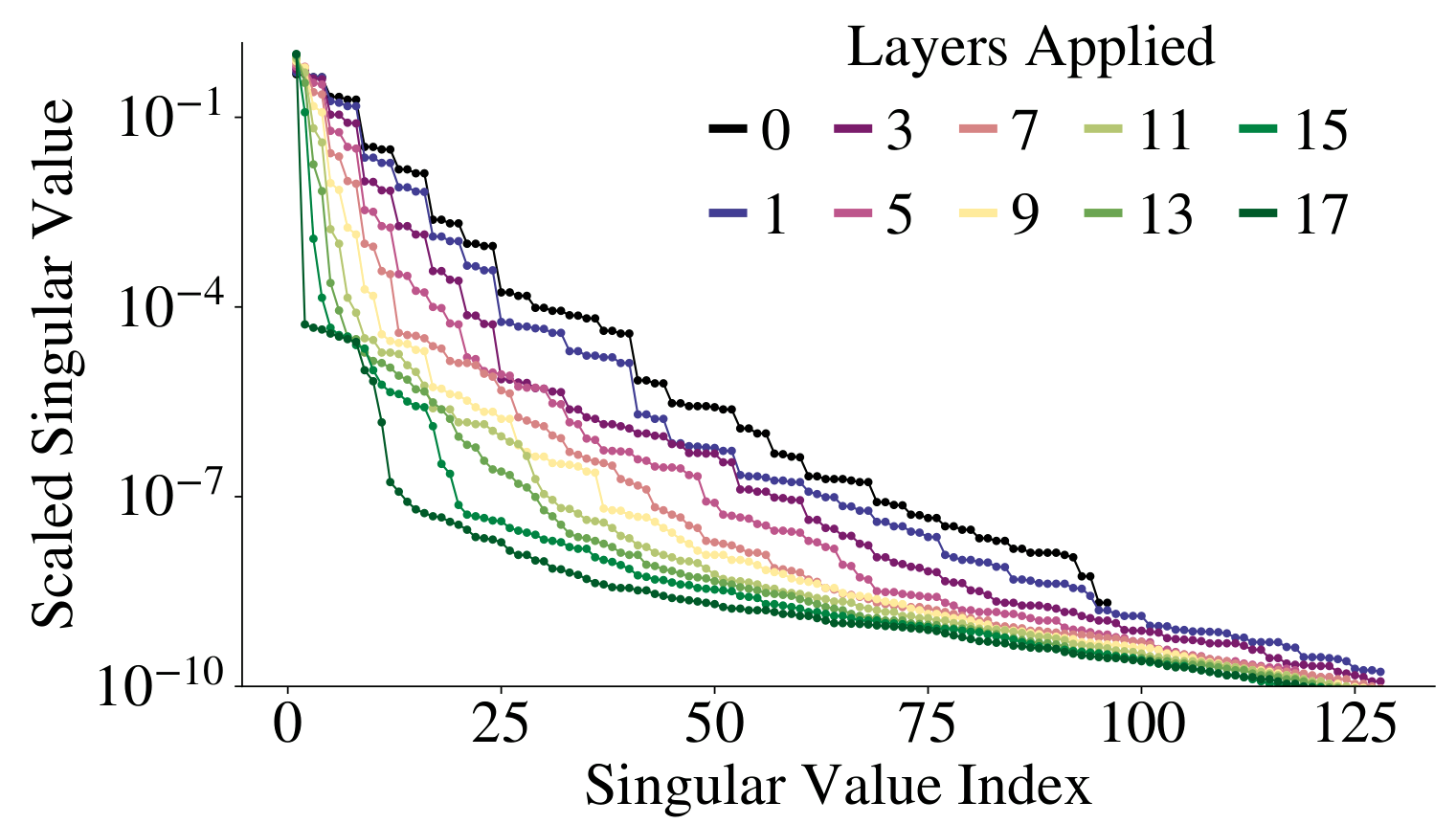}
\vspace{-2mm}
\caption{\textbf{ Faster decay of singular values with increasing layers.} Here we provide numerical results motivating that the compilation is not directly impeded by circuit depth. The target unitary ($V_{\text{Targ}}$) considered is the same MPO trained on in Fig.~\ref{fig:Numerics}a) for the propagator of the 1D Transverse-Field Ising model. We optimize an $L=17$ layer circuit to maximize the overlap with $V_{\text{Targ}}$ then extract the operator-Schmidt coefficients of the resulting MPO after inverse layers of the optimized ansatz are progressively applied, see text for details. The singular value vector is scaled by dividing by the norm. Specifically, we consider the MPOs of the form $(U^{L-i:L})^{\dagger}V_{\text{Targ}}$ for increasing values of $i$. As the data clearly shows, the more layers are applied, the faster the singular values decay, indicating that the dominant cost of the algorithm depends on the bond dimension of the target unitary, rather than the number of layers in the parameterized circuit. }
\label{fig:SingularValues}
\end{figure}

In Sec.~\ref{sec:CreateMPO} we described how the longest time propagator is determined for a maximum allowed bond dimension before the compression of the MPO (the bond dimension is set to 128 in this study).  We found that for the majority of our results, using the same maximum bond dimension was sufficient for the whole optimization to proceed successfully. Only for the most extreme depth circuits, e.g. greater than 60 layers, we found that the optimization would proceed for a large number of iterations, but later may reach a point where the cost function fails to monotonically decrease.
This is due to accumulating truncation errors too great in comparison to the small cost function values, causing the gate environments to become unfaithful. For these restricted cases, we found that restarting the optimization with a modest increase in training bond dimension e.g. from 128 to 156, was sufficient to return the optimization to a smooth decrease of the cost function.


The simulation of quantum circuits with tensor networks is typically expected to require an exponentially growing bond dimension as a function of circuit depth. We stress that for our application, by the efficient use of environment tensors, and the circuit layers effectively acting as disentanglers, the computational cost is instead dominated by the bond dimension of the target unitary.
We observe that training on a comparable bond dimension to the target MPO is sufficient to achieve accurate compilation. Those results are corroborated by an observation that optimizing over deep circuit does not significantly increase the required bond dimension to represent the environment MPOs.

We demonstrate this point with numerical results in Fig~\ref{fig:SingularValues}. We take the target MPO for the 1D Transverse-Field Ising model in Fig.~\ref{fig:Numerics}a) and optimize an $L=17$ layer circuit, $U$, to maximize the overlap with $V_\text{Targ}$.
After progressively applying inverse layers of the parameterized circuit to $V_\text{Targ}$, we study the rate of decay of the operator-Schmidt coefficients. Specifically, for the MPO $(U^{L-i:L})^{\dagger}V_{\text{Targ}}$, we move the orthogonality center onto the central MPO site tensor, contract this with its neighboring tensor, and SVD the joint tensor to inspect the singular values after normalization.
In Fig.~\ref{fig:SingularValues}, we show these normalized singular values, and it is clear that as more inverse layers are applied with increasing $i$, the faster the singular values decay. This behavior can be expected, because as the optimization progresses and the compilation error decreases, the product $(U^{L-i:L})^{\dagger}V_{\text{Targ}}$ moves closer to the identity MPO. This behavior of our circuit layers acting as disentanglers is key for our compilation being able to perform deep circuit optimizations.

\section{Discussion}\label{sec:discussion}

Simulating quantum dynamics is a leading application for achieving a quantum advantage on near-term quantum computers, however the high circuit depths remain prohibitive on current NISQ devices. To advance the time to productive use of quantum computers, it is desirable to maximally use classical computation to optimize quantum circuits so the algorithms are run more efficiently. In this direction, we build upon work performing classical pre-computation and optimization of quantum circuits, and present a compression of Trotter unitaries for quantum dynamics. Our MPO-based optimization found error reductions by up to a factor of $10^4$ compared to equivalent depth Trotterizations, or alternately for a target error rate a depth compression up to a factor of 6.4. By the efficient use of environment tensors, our work extends the literature by enabling far deeper circuit compilations than previously possible, optimizing our results and we show the successful optimization of circuits with up to 64 layers of SU(4) gates. These deep circuit optimizations are enabled by the gate layers acting as disentanglers to the target MPO, allowing deep circuits to be targeted beyond prior works. To demonstrate that our approach can flexibly compile to the connectivity of a target quantum computer, we show the compression of a unitary directly onto the 2D IBM Heavy-Hex topology.

While the computation of gate environments can be made efficient in the limit of large qubit numbers and circuit depths, the success of the algorithm remains dependent on a challenging optimization.
We do not claim that the compiled circuits we present are optimal in error versus the propagator for a given circuit depth. Future work could explore more advanced global optimization algorithms e.g. using Riemannian optimization over the unitary manifold~\cite{luchnikov2021riemannian}, however the simple and local gate update used was sufficient to reduce errors compared to Trotterization for every circuit depth tested. This demonstrates practical utility for reducing the resource requirements in the quantum simulation of dynamics. for achieving up to 4 orders of magnitude reduction in approximation errors and over a factor of 6 reduction in circuit depth.

\begin{acknowledgments}

We acknowledge the use of the TensorOperations package \cite{TensorOperations.jl} for our tensor contractions. We thank Faisal Alam for helpful suggestions and contributions to the technical discussion.
J.G. was supported with funding and computational resources provided by AWE.
L.C. was supported by Laboratory Directed Research and Development program of Los Alamos National Laboratory under project number 20230049DR as well as by the U.S. Department of Energy, Office of Science, Office of Advanced Scientific Computing Research under Contract No. DE-AC05-00OR22725 through the Accelerated Research in Quantum Computing Program MACH-Q project.

\end{acknowledgments}

\newpage

\bibliographystyle{quantum}
\bibliography{main.bib}

\clearpage 
\appendix

\setcounter{page}{1}
\renewcommand\thefigure{\thesection\arabic{figure}}
\setcounter{figure}{0} 

\onecolumngrid

\begin{center}
\large{ Supplementary Material for \\ ``Deep Circuit Compression for Quantum Dynamics via Tensor Networks''
}
\end{center}

\section{Hamiltonians}\label{sec:Hamiltonians}

Here we further detail the Hamiltonians studied in our numerical results in Sec.~\ref{sec:results}.

\begin{figure*}[b!]
\centering
\includegraphics[width=0.9\textwidth]{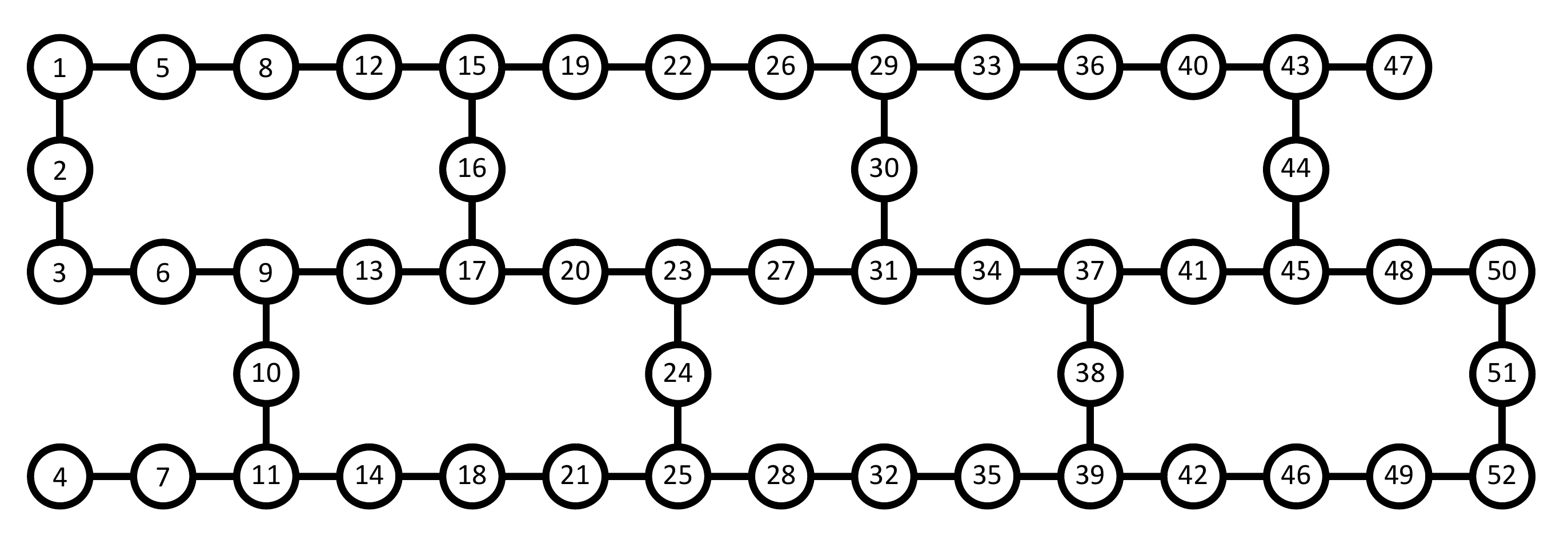}
\vspace{-2mm}
\caption{\textbf{Heavy Hex Topology. }For the 2D compilation demonstration, the Hamiltonian chosen is the Transverse-Field Ising model with the same connectivity graph as the shown 52 qubit subset of the IBM Heavy-Hex topology. To map the local 2D gates onto the MPO, we use a winding across the shortest dimension, as indicated by the qubit numbering.}
\label{fig:HeavyHex}
\vspace{-1.5em}
\end{figure*}

\textbf{1) 1D Transverse-Field Ising model}

\begin{equation}
    H = \sum_{i=1}^{n-1} Z_i Z_{i+1} + h\sum_{i=1}^n X_i
\end{equation}

We target a 200 qubit chain, with a field strength of $h=1.0$ for a total evolution time of $t=2.0$.

\bigskip

\textbf{2) 1D Hubbard model}

\begin{equation}
\begin{split}
    H = -t_{\text{hop}}\sum_{i,\sigma} \left( c^{\dagger}_{i,\sigma}c_{i+1,\sigma} + c^{\dagger}_{i+1,\sigma}c_{i,\sigma} \right) + U\sum_{i} n_{i,\uparrow}n_{i,\downarrow}
\end{split}
\end{equation}

We use a staggered representation where the spin up/down orbitals are neighboring on a 1D chain. We target the Hamiltonian with 50 sites (corresponding to 100 qubits), with $U=4.0$, $t_{\text{hop}}=1.0$, and total evolution time $t=0.2$. We assume access to a 1D chain of qubits with nearest-neighbor interactions; the next-nearest neighbor interaction for the hopping terms therefore requires fermionic SWAP gates in the Trotter unitaries.

\bigskip

\textbf{3) 1D $J_1-J_2$ Model}

\begin{equation}
\begin{split}
    H = J_1\sum_{i=1}^{n-1} \bar{\sigma}_i \cdot \bar{\sigma}_{i+1} + J_2\sum_{i=1}^{n-2} \bar{\sigma}_i \cdot \bar{\sigma}_{i+2} 
\end{split}
\end{equation}

We target a chain of 40 qubits for a total evolution time of $t=1.0$, with $J_1=1.0$ and $J_2 = 0.25$.
We assume access to a 1D chain of qubits with nearest-neighbor interactions; the next-nearest neighbor interaction therefore requires SWAP gates in the Trotter unitaries.

\bigskip

\textbf{4) Transverse-Field Ising model on the Heavy Hex topology}

We also compile circuits for the Transverse-Field Ising model on the 2D Heavy-Hex topology found on IBM superconducting chips. For the 2D compilation, we compile the circuit directly onto the same topology, generating a circuit that has the same local 2D interactions as the real quantum device. We target the 52 qubit subset of qubits shown in Fig.~\ref{fig:HeavyHex}. We set $h=0.75$, and $t=0.5$.

\clearpage

\end{document}